\documentclass[aps,preprint,showkeys,,floatfix]{revtex4-1}
\usepackage{amsmath}
\usepackage{latexsym}
\usepackage{float}
\usepackage{amssymb}
\usepackage{graphicx}
\usepackage{epsfig}

\newcommand{\theeq}{\theta_\mathrm{eq}}
\newcommand{\theeqw}{\theta_\mathrm{eq,wall}}

\begin{document}

\title{
Modelling receding contact lines on superhydrophobic surfaces
}

\author{B.\ M.\ Mognetti  and J.\ M.\ Yeomans,\\
The Rudolf Peierls Centre for Theoretical Physics,\\
 1 Keble Road, Oxford, OX1 3NP, United Kingdom.\\
}

\begin{abstract}
 We use mesoscale simulations to study the depinning of a receding contact line on a superhydrophobic surface patterned by a regular array of posts. In order that the simulations are feasible, we introduce a novel geometry where a column of liquid dewets a capillary bounded by a superhydrophobic plane which faces a smooth hydrophilic wall of variable contact angle. We present results for the dependence of the depinning angle on the shape and spacing of the posts, and discuss the form of the meniscus at depinning. We find, in agreement with \cite{Cohen}, that the local post concentration is a primary factor in controlling the depinning angle, and show that the numerical results agree well with recent experiments. We also present two examples of metastable pinned configurations where the posts are partially wet. 

\end{abstract}

\maketitle

electronic mail: b.mognetti1@physics.ox.ac.uk; j.yeomans1@physics.ox.ac.uk

\section{Introduction}

Hydrophobic surfaces patterned by micron-scale posts become superhydrophobic \cite{RSN-08}. Drops resting on the posts, in the Cassie or fakir state \cite{Cassie}, can have contact angles approaching $180^\circ$, and roll surprisingly easily. An increasing number of actual or potential applications for superhydrophobic surfaces are under investigation: these include energy efficient condensers \cite{Cond}, water harvesting devices \cite{Harv}, and stay dry surfaces \cite{RSN-08}.

Contact angle hysteresis is a term describing the difference in the advancing and receding contact angle of a drop as it moves across a surface \cite{Q-book}. In general the hysteresis depends on both the drop velocity and on its shape which is, in turn, determined by how it is driven. Viewed as surfaces with well controlled roughness, superhydrophobic substrates provide an interesting model system for investigating contact angle hysteresis. Moreover a better understanding of hysteresis on such surfaces may be useful in their development as devices. For surfaces patterned with posts the advancing contact angle is $\approx 180^\circ$ \cite{SqPosts} and the fluid motion is belived to be controlled by the receding contact line \cite{Joanny,RQ-2009}. Therefore in this article we present numerical results describing the motion of a contact line as it recedes across a superhydrophobic surface. In particular we are interested in how the geometry of the posts determines the receding contact angle at which the drop starts to move.

Several authors have discussed how different lattice geometries and post shapes can affect hysteresis on superhydrophobic surfaces \cite{SHfractal,SHmultiscale,SHshape,SHtortuoseshape,HysSP,Hys1,Hys2,Hys3,Cohen,RQ-2009,Joanny,SqPosts,hole2,HalimLang,Collapse1}. The importance of the deformation of the triple line \cite{Joanny,RQ-2009,SHtortuoseshape,hole2} has been highlighted, although a full understanding is so far lacking. There are recent experimental results exploring how a contact line depins from a superhydrophobic surface \cite{SqPosts,RQ-2009}.

We aim to present simulations of contact line depinning to compare to the experimental results and theoretical approaches. The fluid is described by a Ginzburg-Landau free energy model and the hydrodynamics by the Navier-Stokes equations. Hence our work is relevant to length scales above $\sim 50$nm where fluctuations are not dominant and, because we neglect gravity, drops smaller than the capillary length $\sim 1mm$.
Such mesoscale simulations of three-dimensional drops on posts are very demanding and we found that it was not possible to use enough posts beneath the drop to obtain satisfactory results because of unphysical dependences on the initial position of the drop and on inertial effects. Therefore we describe a novel capillary dewetting geometry, where one wall of a microchannel is superhydropobic. The facing wall is smooth, and has a variable contact angle which can be tuned to drive depinning.

This is described in the next section, together with details of the model fluid that we will consider. Then, in Sec.~\ref{res_post} we present results for the depinning angle on several post geometries and, in Sec.~\ref{MenDef}, interpret these in terms of the interface shape as it depins. Choi {\em et al.} \cite{Cohen} have recently proposed a phenomenological description of the dependence of depinning angle on geometry. In Sec.~\ref{res_cohen} we show that this is usually a very good representation of the data and we point out when, and why, it can fail. In Sec.~\ref{Comparison} we show that the simulations agree well with experiment. Finally, in Sec.~\ref{Conclusion} we summarise and discuss our results. 

\begin{figure}
\includegraphics[angle=-90,scale=0.4]{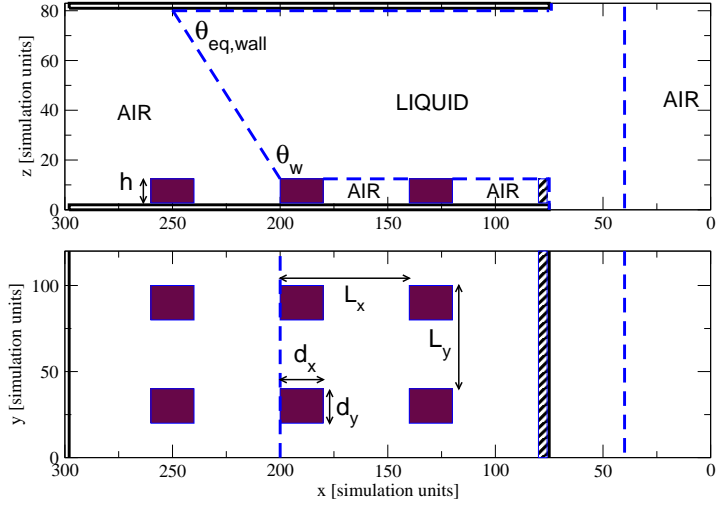} \\
(a)\\
\includegraphics[angle=0,scale=0.4]{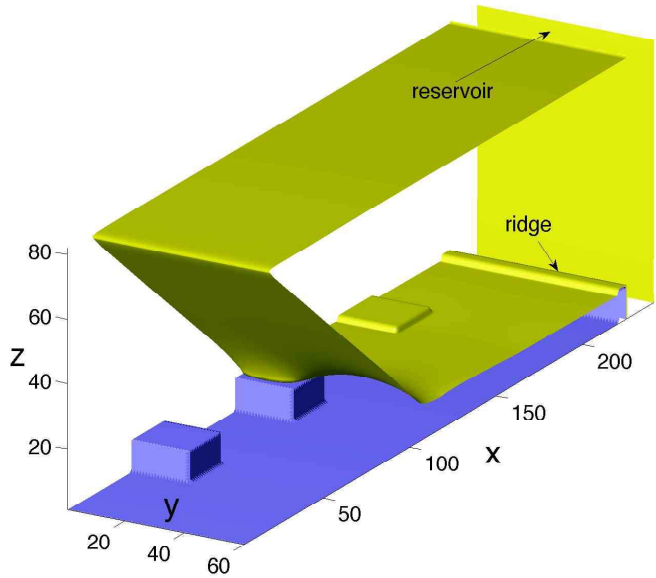}\\
(b)
\caption{
(a) Side and top view of the simulation geometry. A superhydrophobic surface patterned by square posts with $\theeq=100^\circ$ \cite{RQ-2009} faces a hydrophilic smooth channel with contact angle $\theeqw$. A slab of liquid lies between the top of the posts and the wall opposite and is terminated in a liquid reservoir (on the right in the Figure). A solid ridge  separates the channel from the reservoir. We use periodic boundary conditions in both the $x$ and $y$ directions and, in the reservoir, also in the $z$ direction. (b) Three dimensional view of the channel.
}\label{FigGeom}
\end{figure}

\section{Model and Method}

\subsection{Simulation Geometry}\label{SecSystem}

The system we investigate is shown in Fig.\ \ref{FigGeom}. A channel is formed by two planes lying perpendicular to the $z$-axis at $z=0$ and $z=L_z$. The channel is filled with coexisting liquid and gas such that an interface between the two phases lies within the channel. In the simulations the interface moves in the $x$-direction and periodic boundary conditions are imposed along $y$.

Each end of the channel terminates in a reservoir of fluid and we take periodic boundary conditions along $x$ (such that a second interface lies within the reservoir). This equalises the pressure of the gas reservoir and the gas in the channel. In the reservoir periodic boundary conditions are applied along both $y$ and $z$. This produces a flat meniscus which does not exert a Laplace pressure on the fluid in the channel. A similar geometry (but with chemical pinning) has recently been used as a valuable tool to measure the equilibrium contact angle in molecular dynamics simulations \cite{Binder2010}.

The bottom surface of the channel is superhydrophobic. This is achieved by patterning the plane by posts, of height $h=12$ and rectangular cross section ($d_x$, $d_y$), and choosing a hydrophobic equilibrium contact angle $\theeq=100^\circ$ \cite{RQ-2009,SqPosts}. The spacing between the posts is $s_x-d_x$ and $s_y-d_y$ along $x$ and $y$ respectively so that the fraction of surface covered by the posts is $\phi=d_x d_y/(s_x s_y)$. We choose $s_x =s_y =60$ and vary $d_x$ and $d_y$ to measure the effect of post concentration on the depinning. Between the superhydrophobic substrate and the reservoir we place a ridge of the same height as the posts, $h$, but which extends across the channel (see Fig.\ \ref{FigGeom}). This avoids the water in the reservoir wetting the base of the superhydrophobic surface (to give a Wenzel state \cite{Wenzel}).

The surface at $z=L_z$ is chosen to be smooth and hydrophilic, with a variable equilibrium contact angle $\theeqw$.  We take $L_z$ between $50$ and $80$ to ensure that the liquid-gas interface is planar in the vicinity of this wall. For high values of $\theeqw$ the liquid dewets the channel. We define the apparent contact angle that the interface makes with the superhydrophobic surface as 
\begin{eqnarray}
\theta_W = 180^\circ - \theeqw 
\end{eqnarray}
(see Fig.\ \ref{FigGeom}). Note that $\theta_W$ is measured sufficiently far from the triple line that the liquid--gas profile has become flat. The receding angle $\theta_R$ is then the value of $\theta_W$ at which depinning takes place.

The size of the simulation box along $z$ is $L_z$. Along $y$ the periodic boundary conditions allow a box of width $L_y = s_y$. At high post concentration $\phi$ it is sufficient to consider two rows of posts along $x$, such that the front is pinned to the second, in the direction of decreasing $x$. This gives a simulation box size along $x$, $L_x \sim 200$. At smaller post concentrations we observed that $\theta_R$ could be affected by the bounding ridge. Therefore a longer system was needed with three rows of posts, as shown in Fig.\ \ref{FigGeom}, corresponding to $L_x \sim 300$. 

We turned to using the channel geometry to investigate the receding contact line after attempting to perform simulations with cylindrical drops lying on square posts and driven by a body force. These proved very difficult because limitations in system size meant that only a few posts could be used in the direction of motion of the drops. Inertial effects as the drop depinned led to results strongly dependent on the initial position of the drop with respect to the underlying posts.

\subsection{The model and numerical details}\label{SecModel}
We use a binary fluid model (with components $A$ and $B$) described by a Landau  free energy $\Psi$ written in terms of the total density distribution $n=n_A+n_B$ and the order parameter $\varphi=n_A-n_B$ \cite{Swift,BriantBinary}
\begin{eqnarray}
\Psi &=& \int_V (\psi_b + \frac{\kappa}{2} (\partial_{\alpha}\varphi)^2) \;dV - \int_S h \cdot \varphi \, dS  
\label{freeen0}
\end{eqnarray}
where $V$ is the volume and $S$ the surface of the computational domain. The bulk contribution to the free energy density
\begin{eqnarray}
\psi_b &=& \frac{c^2}{3} n \ln n + A \left( -\tfrac{1}{2} \varphi^2 + \tfrac{1}{4} \varphi^4 \right)
\label{freeen}
\end{eqnarray}
includes an ideal term in $n$ which controls the compressibility of the fluid and a quartic contribution in $\varphi$ which leads to two coexisting bulk phases with $\varphi= \pm 1$. The gradient term in Eq.\ \eqref{freeen} accounts for the excess free energy associated with an interface. It is related to the surface tension between the two phases by $\gamma = \sqrt{8\kappa A/9}$  and to the interface width through $\xi = \sqrt{\kappa/A}$  \cite{BriantBinary}. We use $A=0.02$ and $\kappa=0.04$ which give $\xi=1.41$ and $\gamma=0.0267$. The surface term controls the equilibrium contact angle $\theeq$ of the fluid on the solid substrate \cite{Cahn}
\begin{eqnarray}
h &=&  \sqrt{2 \kappa A} \cdot \mathrm{sign}\Bigg( {\pi\over 2} - \theeq \Bigg) \cdot \sqrt{\cos \Big({\alpha \over 3}\Big)\Big[ 1-\cos \Big( {\alpha\over 3} \Big)\Big]} 
\, ,
\end{eqnarray}
where $\alpha=\cos^{-1}(\sin^2\theeq)$. The functional (\ref{freeen0}) is discretised on a cubic lattice with lattice spacing $\Delta{}x$, and $\Delta{}t$ is the simulation time step. The lattice velocity is then defined as $c=\Delta{}x/\Delta{}t$.

The hydrodynamics of the binary fluid is given by the continuity equation (\ref{eqCon}), the Navier-Stokes equation (\ref{eqNS}) and the convection-diffusion equation (\ref{eqCD})
\begin{eqnarray}
&\partial_{t}n+\partial_{\alpha}(nv_{\alpha})=0 \, , \label{eqCon} \\
&\partial_{t}(nv_{\alpha})+\partial_{\beta}(nv_{\alpha}v_{\beta}) = - \partial_{\beta}P_{\alpha\beta}+ \partial_{\beta}[\eta (\partial_{\beta}v_{\alpha} + \partial_{\alpha}v_{\beta} + \delta_{\alpha\beta} \partial_{\gamma} v_{\gamma}) ]  \, , \label{eqNS} \\ 
&\partial_t{\varphi} + \partial_\alpha \left( \varphi v_\alpha \right) = M \nabla^2 \mu \, , 
\label{eqCD}
\end{eqnarray}
where we have introduced the flow velocity $v$ and the dynamic viscosity $\eta$. In Eq.\ (\ref{eqCD}) the parameter $M$ is a mobility which controls the diffusion of the triple line when the system is pushed out of equilibrium. This is the mechanism by which the contact line moves although no slip boundary conditions ($v=0$) are implemented at the solid wall. The pressure tensor $P_{\alpha\beta}$ in Eq.~(\ref{eqNS}) and the chemical potential $\mu$ in Eq.~(\ref{eqCD}) are derived from the free energy (\ref{freeen0}) and are given by 
\cite{BriantBinary}
\begin{eqnarray}
&\mu  = A\left(- \varphi + \varphi^3 \right)- \kappa \nabla^2 \varphi, \label{eqChemPot} \\
&P_{\alpha \beta} =  \left(p_b - \frac{\kappa}{2}(\partial_\gamma \varphi)^2 - \kappa \varphi \partial_{\gamma\gamma} \varphi \right) \delta_{\alpha \beta} + \kappa (\partial_{\alpha} \varphi)  (\partial_{\beta} \varphi) \, , \label{eqPressTen} \\
& p_b =  \tfrac{c^2}{3}n  +  A \left( -\tfrac{1}{2} \varphi^2 + \tfrac{3}{4} \varphi^4 \right). \nonumber
\end{eqnarray}
The mesoscopic equations (\ref{eqCon}-\ref{eqCD}) are solved using a lattice Boltzmann algorithm \cite{LB} in the implementation described in detail by Pooley  et.\ al.\ \cite{PooleyMRT} which has been shown to be successful in reducing spurious velocities related to lattice artifacts. For further details we refer to Refs.\ \cite{Swift,BriantBinary,PooleyMRT,Pooley-09}. Although our results have been obtained using a binary model, we will be working in a quasi-static regime where a liquid-gas formalism gives the same results. 
We have chosen to use this approach because our experience is that it relaxes efficiently to stable and metastable minima of the free energy and because it can very naturally be extended to dynamical problems when the drop is moving across the surface. Surface Evolver \cite{SurfEv} provides an alternative approach in the static limit \cite{SurfEv1,SurfEv2,SurfEv3}.


\begin{figure}
\includegraphics[angle=0,scale=0.5]{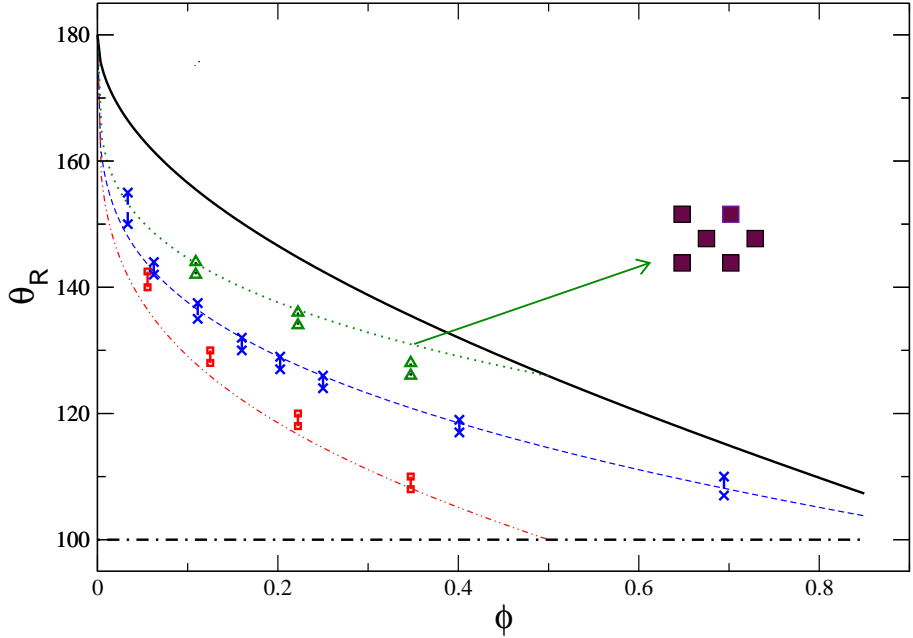}
\vspace{0.1cm}
\caption{
Dependence of the receding contact angle $\theta_R$ on concentration. We consider three geometries: square posts arranged on a square lattice ($\times$), rectangular posts with $d_y=2 d_x$ on a square lattice ($\square$), and square posts on a square centred lattice ($\triangle$). In each case $\theta_R$ lies between the two values shown. For the larger/smaller of these the front dewets/remains pinned. The light lines are the scaling predictions of Ref.\ \cite{Cohen}. The heavy lines refer to ridges and grooves along $x$ ($\theta_R=\theta_C$, full line) and $y$ ($\theta_R=\theeq$, dashed-dotted line).
}\label{FigThetr}
\end{figure}

\section{Results}\label{SecResults}

Fig.\ \ref{FigThetr} summarises the simulation results for the depinning angle $\theta_R$ for several different post geometries.  After discussing the data, we relate it to the shape of the meniscus as the interface is pulled across the posts. We then interpret the results in terms of a phenomenological model of depinning, and compare to experiments on drops on superhydrophobic surfaces.

\subsection{Post geometry}\label{res_post}

{\bf Ridges across the channel.} Before considering a substrate patterned with a more complex post geometry it is instructive to sketch out the bounding cases of ridges and grooves either running across ($s_y=d_y$ in Fig.\ \ref{FigGeom}) or along the channel ($s_x= d_x$). In the first case the system is translationally invariant along $y$ and, noting that the interface is flat, $\theta_R$ follows easily from the Gibbs' criterion \cite{Gibbs}. This states that dewetting occurs if the angle between the meniscus and the wet, horizontal face of the ridge 
is smaller than the equilibrium contact angle of the flat surface. Hence $\theta_R=\theeq$ independent of $\phi$. The simulations reproduce this limit, shown by the dash-dotted line in Fig.\ \ref{FigThetr}. 

{\bf Ridges along the channel.} The case of grooves and ridges along $x$, 
parallel to the flow direction,  is slightly more involved because the liquid-gas interface is not planar. In this case there is no pinning and the channel is spontaneously wet or dewet for $\theta_W>\theta_R$ or $\theta_W<\theta_R$ respectively, while for $\theta_W=\theta_R$ the liquid--gas interface remains stationary. The shape of the interface is independent of its position in the channel and therefore, withdrawing the interface through $\delta x$, the change in free energy is
\begin{equation}
\delta {\cal F} \propto -\gamma\cdot \delta x\cdot \cos \theta_W + \gamma\cdot \delta x \cdot (\phi \cdot \cos \theeq +\phi-1).
\end{equation}
At the threshold angle $\delta {\cal F} = 0 $. Hence $\theta_R = \theta_C$ where $\cos \theta_C=\phi\cos\theeq+\phi-1$ and $\theta_C$ can be identified as the Cassie angle of the superhydrophobic plate as expected \cite{Cassie}. This limit is shown by the full line in Fig.\ \ref{FigThetr}.

{\bf Square posts.} We now consider a bottom plate patterned with square posts. We vary the post size, $d_x = d_y$ while keeping the period of the pattern, $s_x$ and $s_y$, fixed. Fig.\ \ref{FigThetr} shows numerical estimates of the receding angle  $\theta_R$ for different post concentrations $\phi$. For a given $\phi$ we ran simulations at different $\theta_W$. In Fig.\ \ref{FigThetr} we bound $\theta_R$ with  the two values of $\theta_W$ between which the crossover from pinning to dewetting was observed. The results show that $\theta_R$ tends to $\theeq$  for $\phi\to 1$, and to $180^\circ$ with a divergent slope for $\phi\to 0$. Thus the numerical results for $\theta_R$ are, as expected, bounded by the depinning lines of the grooved geometries. 

{\bf Rectangular posts.} 
In Fig.\ \ref{FigThetr}, symbols ($\square$) indicate simulation results for rectangular posts with $d_y=2\cdot d_x$ arranged on a square lattice. As expected depinning is harder than for square posts for a given concentration, primarily because the spacing between posts along $y$ is smaller and the geometry interpolates between squares and ridges along $y$. As $\phi\to 0.5$, $\theta_R \to \theeq$, and continuous ridges form across the channel. Similarly, rectangular posts elongated along the $x$ direction will make depinning easier at a given concentration.

{\bf Face-centred square patterning.} 
Finally we present results for a different arrangement of posts, square posts arranged on a face-centred square lattice. This geometry is shown in Fig.\ \ref{FigThetr}, where the triangles depict the corresponding simulation data for the depinning angle. For a given $\phi$ it is easier for posts on a face-centred square lattice to depin than posts in a square array because of the larger distance between the posts. However, for a given post spacing, the opposite holds true because the post in the middle of the unit cell helps to pin the receding interface. This effect, which becomes dominant for the largest concentration reported in Fig.\ \ref{FigThetr}, will be discussed in more detail in Sec~\ref{MenDef}.

\begin{figure}
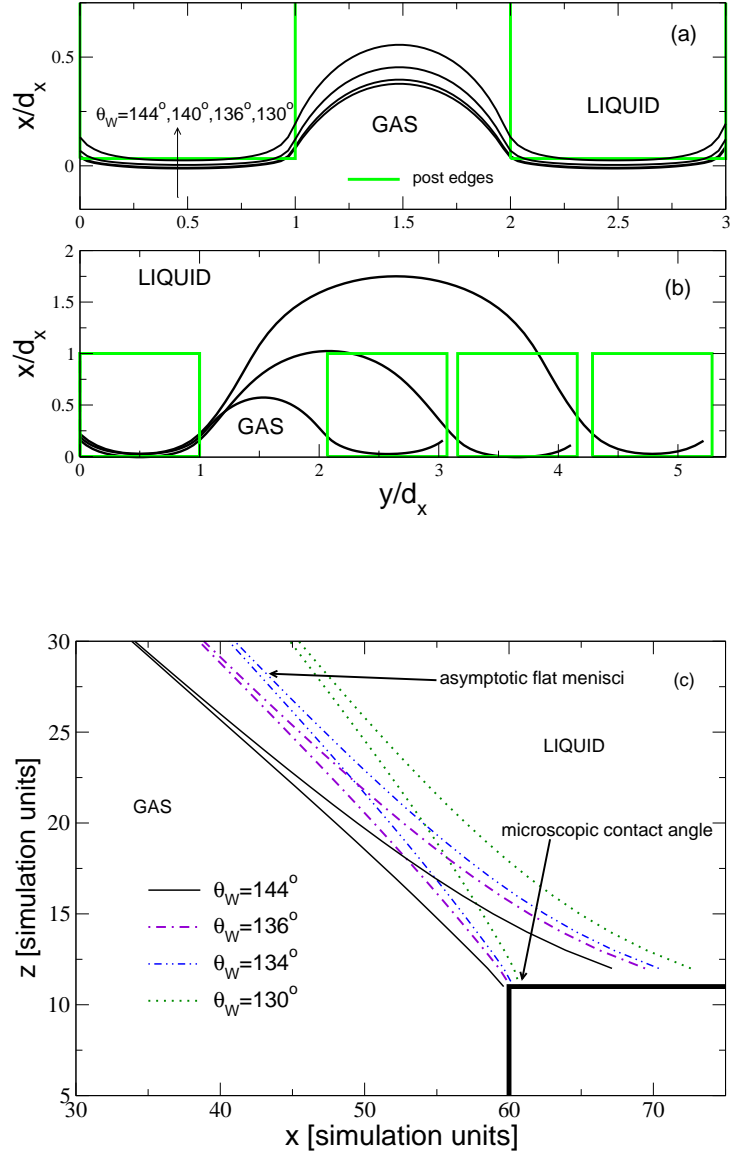

\includegraphics[angle=0,scale=0.4]{fig3a.eps}
\\ \vspace{1.5cm}
\includegraphics[angle=0,scale=0.4]{fig3b.eps}
\caption{Pinned menisci recorded one liquid layer above the posts in the $xy$ plane at (a) constant post concentration $\phi$ for different values of the apparent contact angle  $\theta_W$ and (b) at $\theta_W\approx\theta_R$ for different values of $\phi$. (c) Pinned menisci in the $xz$ plane, comparing cross sections which run through the centre of the posts, and half-way between the posts for different $\theta_W$ and $\phi=0.17$. (The pinned interfaces do not appear to lie exactly at the corners of the posts because we record their position one lattice spacing above the surface and because of the diffuse nature of the interface.)
}\label{FigMeniscus}
\end{figure}
\subsection{Menisci deformations}\label{MenDef}

In general on patterned surfaces the receding contact angle is larger than the Young angle because of the deformation of the liquid-gas interface near the triple line. As the liquid bulges between two posts, as shown in Fig.\ \ref{FigMeniscus}, the interface in the vicinity of the post corners is pulled forwards thus decreasing the local contact angle and facilitating depinning. This deformation is strongly affected by details of the patterning geometry. It occurs in the vicinity of the posts, and the channel height $L_z$ is chosen to be sufficiently large that the interface is essentially planar near the top of the channel. This ensures that the depinning angles are independent of $L_z$ \cite{CapFill}.

In Fig.\ \ref{FigMeniscus}(a) and (b) we show the profiles of pinned menisci in the $xy$ plane one lattice spacing above the top of the posts. The first of these plots shows that, as $\theta_R$ is decreased, the meniscus remains pinned on the back of the posts, but is gradually pulled forwards between the posts. Depinning occurs at $\theta_R = 125^\circ$. (In the Figure the pinned interface does not appear to lie exactly at the post edges because the interface is diffuse.)

Fig. \ref{FigMeniscus}(b) shows how the deformation between the posts depends on different values of the concentration $\phi$ for $\theta_W \approx \theta_R$. As $\phi$ decreases the larger gaps mean that it is easier for the interface to deform, leading to a higher $\theta_R$. Moreover the interface shape become increasingly $cosh$-like, in agreement with \cite{RQ-2009,Joanny}.

In Fig.~\ref{FigMeniscus}(c) we compare the shape of the menicus in the $xz$ plane, taking two cross sections, one cutting the centre of a post, and one lying halfway between two rows of posts. As $\theta_W \rightarrow \theta_R$ the difference between the two profiles increases, meaning a bigger distortion in the vicinity of the posts in agreement with Fig.\ \ref{FigMeniscus}(b). Notice that, because the interface in the reservoir is flat (Fig.\ \ref{FigGeom}), the Laplace pressure is zero and the pinned liquid--gas meniscus has zero mean curvature. (This is consistent with drop hysteresis experiments \cite{RQ-2009} where the radius of the drop is much bigger than the length scale of the surface patterning.) Therefore, as the liquid bulges forward, the $xy$ profile of the interface between the posts becomes convex (to the liquid) and hence, to preserve the condition of overall zero curvature, the $xy$ profile in the centre of the posts must become concave, as shown in Fig.\ \ref{FigMeniscus}(a). As this curvature increases the contact angle with the top of the post decreases until it reaches $\theeq$ when depinning can occur.

\begin{figure}
\includegraphics[angle=0,scale=0.45]{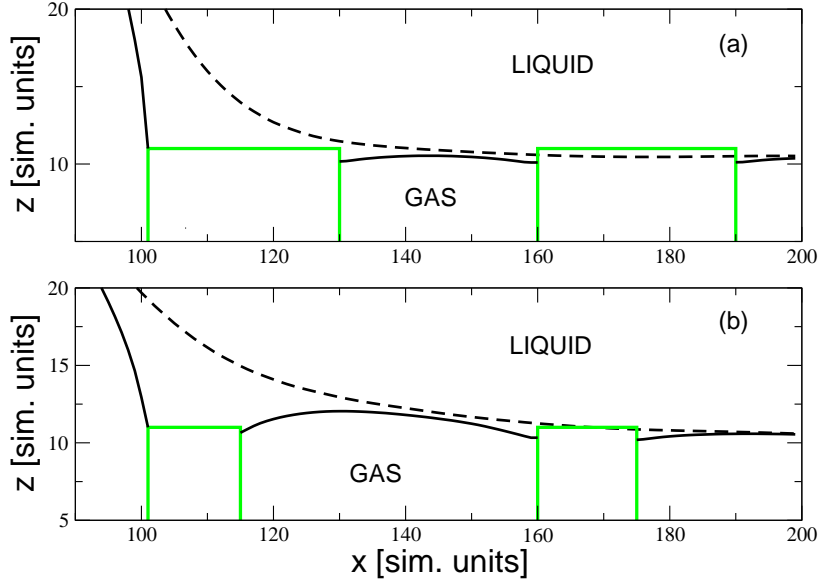}
\caption{
Meniscus profile in the $xz$ plane showing that the curvature under the liquid between the first and second rows of posts increases with decreasing concentration: (a) $\phi=0.25$ (b) $\phi=0.0625$. The broken line shows the interface profile halfway between the posts and the full line the profile in the centre of the posts. (The pinned interfaces do not appear to lie exactly at the top corners of the posts because of the diffuse nature of the interface).
}\label{FigBridge}
\end{figure}
As the post concentration is decreased the meniscus {\em underneath} the liquid does not remain flat but increasingly bulges between the first and second row of posts as shown in Fig.\ \ref{FigBridge}. This occurs because the meniscus is curving sharply upwards to meet the top surface of the channel unless it is pinned by the posts. Hence upwards curvature between  the posts contributes to lowering the interfacial free energy. Recent experiments have shown that small drops can be left behind on the posts as the interface moves forward \cite{neck} and, although we were unable to access a parameter range where this occurred in the simulations, the distortion of the interface under the drop may be a precursor of such a pinch off.

\begin{figure}
\includegraphics[angle=0,scale=0.4]{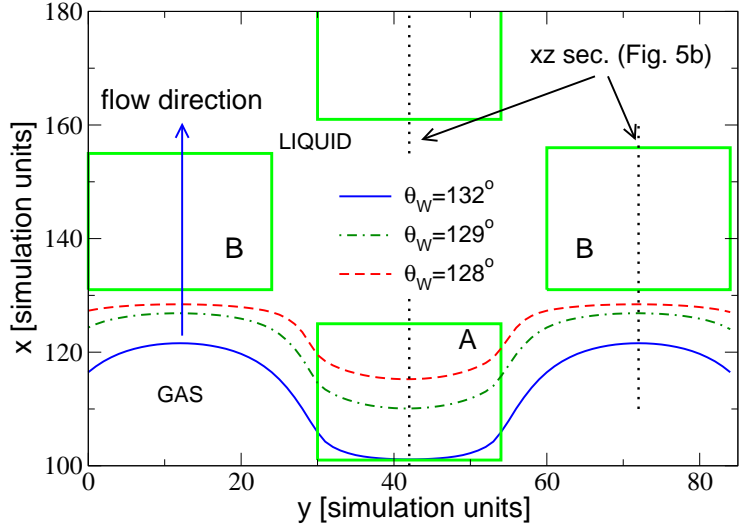} \\ 
\vspace{0.3cm}
(a) \\
\vspace{1.2cm}
\includegraphics[angle=0,scale=0.4]{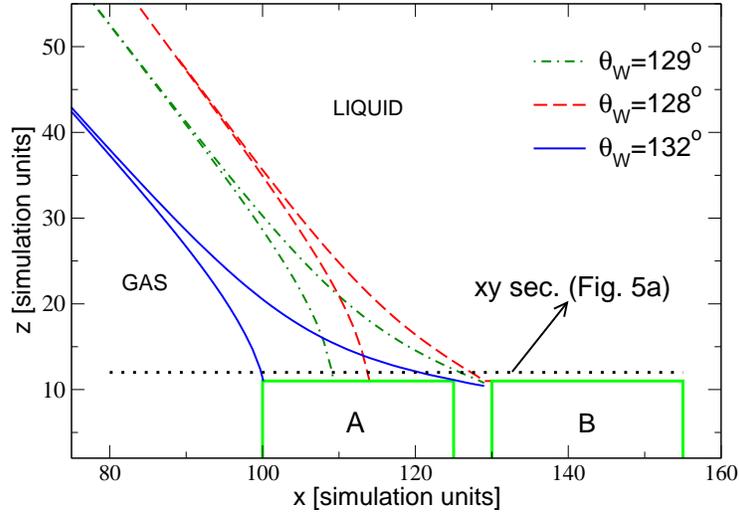} \\
\vspace{0.25cm}
(b)
\caption{
Pinned menisci shapes for the face-centred square patterning for $\phi=0.347$ and three different $\theta_W$. (a) Profiles in the $xy$ plane one lattice spacing above the top of the posts (along the dotted line in Fig.\ \ref{FCC}b). (b) Profiles in the $xz$ plane for two values of $y$ (corresponding to the dotted lines in Fig.\ \ref{FCC}a). For $\theta_W=132^\circ$ the interface is pinned to the A and B posts whereas for $\theta_W=129^\circ$ and $\theta_W=128^\circ$ it is pinned to the B post, leaving the A posts partially wet. In (a) the profiles do not, at first sight, appear to be pinned to the post B because they are recorded one lattice node above the posts and the meniscus meets the dry face of B at an angle which approaches 90$^\circ$ as $\theta_W$ increases. This is apparent in frame (b). 
}\label{FCC}
\end{figure}
We now consider the face-centred square arrangement of posts. At low post concentrations, ($\phi=0.109$ and $\phi=0.22$ in Fig.\ \ref{FigThetr}), depinning proceeds as for square posts: once  the interface has completely depinned from the rear edge of a row of posts the channel dewets. For $\phi=0.347$, however, we find that depinning occurs through a two steps process. This is illustrated in Fig.\ \ref{FCC} where we plot pinned profiles recorded in the $xy$ plane and in the $xz$ plane. The profiles in the Figure indicate that for $\theta_W = 132^\circ$ the front remains pinned to the rear edge of a row of posts, as for simple square patterning. The only difference is that the front between two posts is also pinned to the post in the middle of the unit cell (labelled B in Fig.\ \ref{FCC}). However for $128^\circ \leq \theta_W < 132^\circ$ the contact line, instead of dewetting, remains pinned to the rear of the second row of posts (B), leaving the first row of posts (A) only partially wet. Finally for $\theta_W<128^\circ$ the liquid completely dewets the A posts. The channel will then empty as $\theta_W$ is lower than the value needed to drive either transition. Note that in Fig.\ \ref{FigThetr} we give the value of $\theta_R$ corresponding to the second, final depinning transition.

\begin{figure}
\includegraphics[angle=0,scale=0.5]{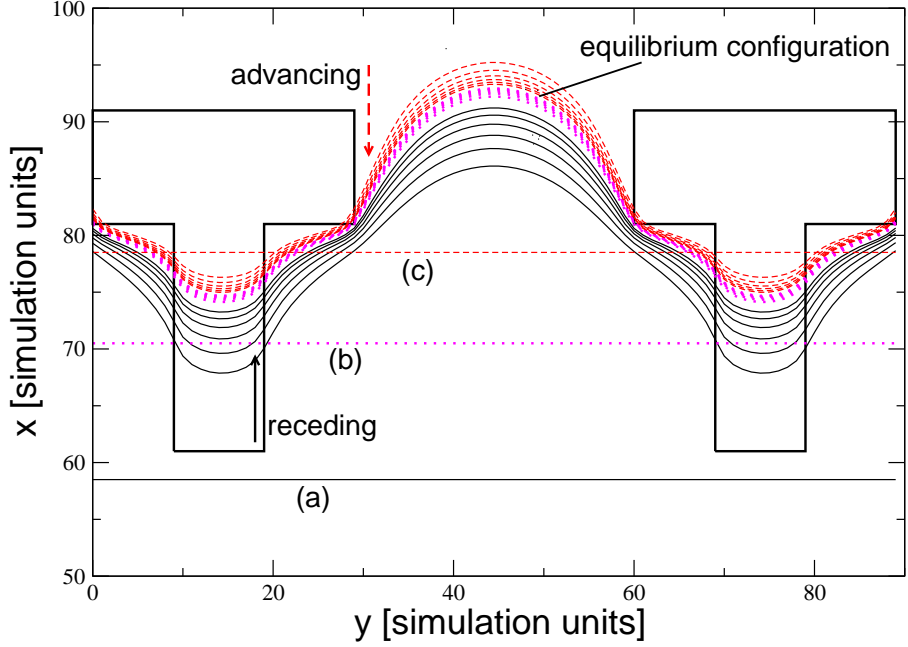}
\vspace{0.25cm}
\caption{ 
Time evolution of the liquid-gas interface on T-shaped posts for $\theta_W=135^\circ$. The profile is recorded one lattice unit above the top of the posts, every $\Delta t=6 \cdot 10^4$ to $39\cdot 10^4$ lattice Boltzmann steps. The starting configurations are the flat interfaces labelled by (a), (b) and (c). The interfaces quickly relax to curved configurations and then much more slowly translate along $x$ to a pinned state in which the posts are partially wet.
}\label{FigMeniscus2}
\end{figure}
The results for square posts shown in Fig.~\ref{FigMeniscus} suggest that the corner of the posts can be dry when there is still pinning on the back edges. However this result could be affected by the diffuse nature of the interface. To demonstrate unequivocally the existence of pinned states with the post tops partially wet we consider T-shaped posts. We chose as initial conditions three flat interfaces lying at different values of $x$, labelled (a) (b) and (c) in Fig.~\ref{FigMeniscus2}, and $\theta_W=135^\circ$. The interface very quickly relaxed to a curved configuation which reflected the presence of the posts. It then advanced or receded much more slowly across the tops of the posts towards a final equilibrium state, lying between the full and dashed profiles in the Figure, that was independent of the starting configuration. The partial depinning occurs because 135$^\circ$ is bounded by the value which depins a $d_y/s_y=10/60$ ridge ($\theta_R=150^\circ$) and that which depins  $d_y\times d_x=30\times 10$ rectangular posts ($\theta_R = 131^\circ$). For more complicated geometries, it is possible that several metastable pinned configurations could exist.

\subsection{An estimate of the depinning threshold}\label{res_cohen}

In a recent paper Choi {\em et al.} \cite{Cohen} proposed a phenomenological theory for the depinning angle of a receding contact line. They argued that the receding contact angle can be estimated using a Cassie-like argument, analogous to that described in Sec.\ \ref{res_post} for a surface patterned by ridges and grooves along $x$, but using the local post concentration $\phi_\ell$ sampled by the interface as it advances a small distance beyond pinning
\begin{equation}
\cos \theta_R= \phi_\ell\cos\theeq +\phi_\ell-1 \, .
\label{coheneq}
\end{equation}
Eq.\ (\ref{coheneq}) assumes that, when the interface depins, it preserves its shape as it moves forwards. Predictions of the scaling relation (\ref{coheneq}), for each of the simulation geometries we consider, are reported in Fig.\ \ref{FigThetr} as light lines.

For square posts the linear density sampled by the interface is $\phi_\ell = 
\sqrt{\phi}$. As is apparent from Fig.\ \ref{FigThetr}, the estimate~(\ref{coheneq}) is in excellent agreement with the numerical results over a wide range of $\phi$. Agreement is essentially exact at high post concentrations, where the meniscus deformation is confined to within the first line of wetted posts.
At low $\phi$, deviations from Eq.~(\ref{coheneq}) start to be apparent, in particular at the lowest post concentration simulated $\phi = 0.0333$. This is because the meniscus is more strongly deformed and probes the region of the channel with no posts. Hence it changes its shape as it moves forwards.

It was not possible to obtain simulation data for lower $\phi$ because this would require too big a simulation box. However, the discrepancies might be expected to be larger for rectangular posts elongated in the $y$ direction which, for a given $\phi$, extend less far along $x$. Choosing, as in Fig.\ \ref{FigThetr}, $d_y = 2 \cdot d_x$ gives $\phi_\ell = \sqrt{2 \phi}$. The agreement is still pleasing, but less close than for square posts. Eq.~(\ref{coheneq}) gives a pinning angle which is slightly too high, and the increase in the discrepancy with  decreasing $\phi$ is clearly seen. This occurs because depinning is facilitated by the detailed shape of the interface between subsequent rows of posts, which is not accounted for the theory.

By contrast, for the face-centred square post geometry the prediction of Eq.\ (\ref{coheneq}) (taking $\phi_\ell = \sqrt{\phi/2}$) underestimates the depinning angle, and becomes much less accurate for large $\phi$. This is because, as argued in Sec.\ \ref{MenDef}, the  post in the middle of the unit cell helps to pin the receding interface.  In particular, for the highest post concentration $\phi = 0.374$, Eq.\ (\ref{coheneq}) predicts a $\theta_R$ which is compatible with the first dewetting transition of Fig.\ \ref{FCC}, rather than reproducing the second depinning which finally dewets the channel.

\begin{figure}
\includegraphics[angle=0,scale=0.5]{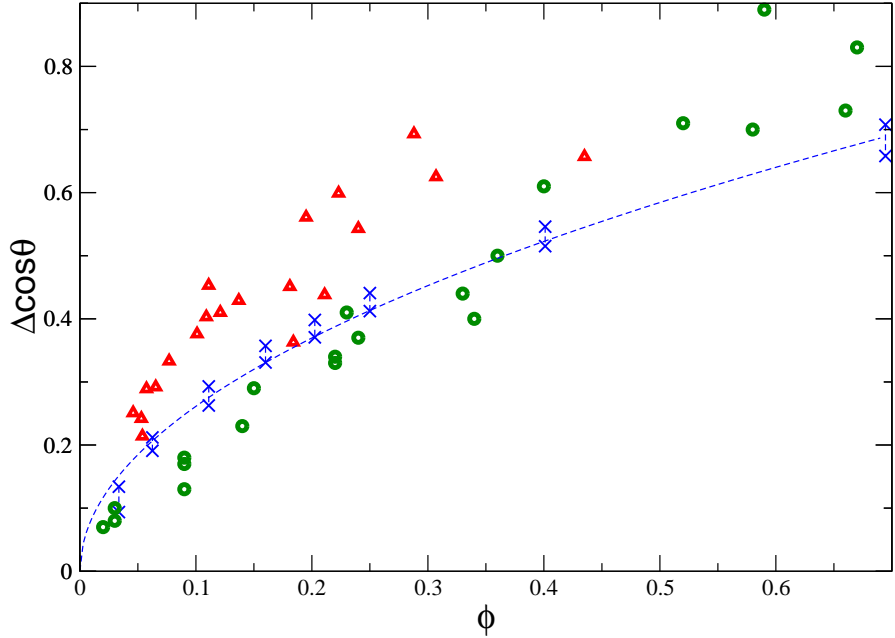}
\caption{
Contact angle hysteresis as a function of the post concentration $\phi$. Circles: experimental results from \cite{SqPosts}. Triangles (digitised): experimental results from \cite{RQ-2009}. Crosses: simulations for a surface patterned with square posts. Broken line: scaling prediction from \cite{Cohen}. Experimental errors are of order the scatter in the data. 
}\label{FigMain}
\end{figure}
\subsection{Comparison to experiment}\label{Comparison}

We next compare the simulation results to experiments measuring drop hysteresis on superhydrophobic surfaces \cite{RQ-2009,SqPosts}. In two independent sets of experiments, carried out by Reyssat and Qu\'er\'e \cite{RQ-2009} and by Priest {\em at al.} \cite{SqPosts}, drops were placed on a superhydrophobic surface which was then tilted. The advancing and receding contact angles, $\theta_A$ and $\theta_R$, were measured at the point where the drop started to move, and the data for each experiment was reported by plotting $\Delta \cos\theta \equiv \cos \theta_A - \cos \theta_R$ as a function of the post concentration. In the experiments described in Ref.~\cite{RQ-2009} the superhydrophobic surface was patterned with circular posts of diameter 2.4 $\mu$m and had an equilibrium contact angle $100^\circ$. The radii of the drops was between $1.3$ and $1.7$ mm. In Ref.~\cite{SqPosts}, by contrast, the posts were square with sides of 20 $\mu$m, the surfaces had an equilibrium contact angle between $95^\circ$ and $115^\circ$ and the drop radii were $< 2$ mm.

There are differences between the experimental protocol and our simulations. Firstly we consider a slab of liquid which is translationally invariant along $y$ sufficiently far from the posts, rather than a drop which forms a spherical cap. Secondly in the experiments the drop is pushed by gravity, a body force, rather than by a capillary force due to the changing contact angle of the upper wall of the microchannel. These should not be important differences for drops of the size considered in the experiments. Thirdly we assume in the simulations that depinning is controlled by the receding contact angle, and we do not measure an advancing contact angle. For the purposes of comparison we set $\theta_A=180^\circ$. Other sensible choices (eg.\ in \cite{SqPosts} $\theta_A$ is measured to lie between $160^\circ$ and $170^\circ$) make little difference to the estimate of $\Delta \cos\theta$.

Fig.\ \ref{FigMain} shows that simulation results compare well with experiments, in particular for the square posts \cite{SqPosts}. Indeed the agreement is surprisingly good given the differences between the experimental and simulation geometries. This supports the assertion that drop hysteresis on superhydrophobic surfaces is controlled by the receding contact line \cite{RQ-2009}.

\section{Conclusion}\label{Conclusion}

We have introduced a novel simulation geometry, a microchannel with a superhydrophobic wall. Considering capillary dewetting of the channel has allowed us to obtain numerical results for the receding contact angle $\theta_R$ of several superhydrophobic surfaces. We used sufficiently high channels that $\theta_R$ only depends on the details of the superhydrophobic surface, and therefore we expect our results to be relevant to drop geometries, as long as the drops are sufficiently large, and as long as depinning is controlled by the receding contact line. Comparison to experiment indicates that this is indeed the case.

Our main results are summarised in Fig.\ \ref{FigThetr} which shows the receding contact angle at depinning as a function of concentration for square and rectangular posts arranged on a simple square or a face-centred square lattice. We find, as first discussed in \cite{Cohen}, that the primary control parameter for depinning is $d_y/s_y$, the fraction of space occupied by the posts in the direction perpendicular to the forcing. This is because depinning is easier if the interface can more easily deform between the posts. Deviations from this scaling manifest themselves once the interface penetrates beyond the posts, for example, for square posts at low concentrations, or if it reaches a second row of posts, as for face-centred square patterning. 

The existence of pinned menisci with partially wet posts has been an open question \cite{PartWetPost}. We find that this is possible for posts arranged in a face-centred square lattice where the interface depins in two steps. We have also identified a metastable pinned state for which T-shaped posts are partially wet. However, detecting pinned menisci where the posts are partially wet in experiments is likely to be challenging due to inertial effects.

From a methodological standpoint, we hope that the geometry we consider here will prove useful in future simulations of depinning. Open questions which we will aim to address include dynamic contact angle hysteresis -- measured when the liquid is moving -- and the response of a moving slab of fluid to a driving force provided by changing the contact angle of the wall. It would be of interest to fabricate an experimental realisation of the patterned channel used in this work, with the contact angle of the hydrophilic flat wall $\theeqw$ controlled by an electrowetting potential. Looking to applications, contact line pinning by posts is being used to control fluid motion in microchannels \cite{pinning} and an understanding of pinning may help design better slippery surfaces, or microchannels with anisotropic filling properties.

\end{document}